\documentclass[doublecol]{epl2}

\usepackage{latexsym}
\usepackage{amsfonts}
\usepackage{amssymb}
\usepackage{amsmath}
\usepackage{graphicx}

\title{Observation of vortices and hidden pseudogap from scanning tunneling spectroscopic studies
of electron-doped cuprate superconductor $\rm Sr_{0.9}La_{0.1}CuO_2$}
\shorttitle{Observation of vortices and hidden pseudogap in $\rm Sr_{0.9}La_{0.1}CuO_2$}

\author{M.~L. Teague\inst{1} \and A.~D. Beyer\inst{1} \and M.~S. Grinolds\inst{1} \and S.~I. Lee\inst{2} \and N.-C. YEH\inst{1}}
\shortauthor{M.~L. Teague \etal}

\institute{
  \inst{1} Department of Physics, California Institute of Technology, Pasadena, CA 91125, USA\\
  \inst{2} National Creative Research Initiative Center for Superconductivity, Department of Physics, Sogang University, Seoul, Korea 121-742
}

\abstract{
We present the first demonstration of vortices in an electron-type cuprate superconductor, the highest $T_c$ (= 43 K) electron-type cuprate $\rm Sr_{0.9}La_{0.1}CuO_2$. Our spatially resolved quasiparticle tunneling spectra reveal a hidden low-energy pseudogap inside the vortex core and unconventional spectral evolution with temperature and magnetic field. These results cannot be easily explained by the scenario of pure superconductivity in the ground state of high-$T_c$ superconductivity.}

\begin{document}

\maketitle

\section{Introduction}
The physical properties of cuprate superconductors exhibit dramatic contrasts between the electron- and hole-type systems~\cite{Yeh05,Yeh07}. Various anomalous phenomena observed in underdoped and optimally doped hole-type cuprates above the superconducting transition $T_c$, such as the low-energy pseudogap (PG) and Fermi arcs occurring at $T_c < T < T^{\ast}$ where $T^{\ast}$ denotes the ``pseudogap'' (PG) temperature~\cite{LeePA06,Fischer07,Damascelli03}, are conspicuously absent in the electron-type~\cite{Yeh05,Yeh07,Kleefisch01,ChenCT02,Matsui05}. Among various theoretical attempts to account for these unconventional and non-universal physical properties of the cuprates, one school of thought is based on the ``one-gap'' or ``preformed pair'' scenario~\cite{LeePA06,Gros06}. This scenario asserts that the onset of Cooper pair formation occurs at $T^{\ast}$ and that the PG state at $T_c < T < T^{\ast}$ is a disordered pairing state with strong superconducting phase fluctuations. In this context, superconductivity below $T_c$ would represent a phase coherent state of the preformed pairs. The other school of thought entertains the possibility of competing orders (COs) coexisting with superconductivity (SC)~\cite{Vojta00,Kivelson03,Demler04,Polkovnikov02} so that both order parameters are responsible for the low-energy quasiparticle excitations. The COs may be established at a temperature different from $T_c$, thereby accounting for the PG phenomena. To date numerous experimental findings~\cite{Krasnov00,LeTacon06,ChenCT07,Beyer08,Boyer07} and quantitative analyses of the experimental data~\cite{Yeh07,ChenCT07,Beyer08,YuBL08} seem to favor the ``two-gap'' scenario. Nonetheless, the debate continues because the phenomenology of coexisting COs and SC does not directly provide a microscopic pairing mechanism that involves the interplay of COs and SC. On the other hand, there is no obvious theoretical reason to preclude the occurrence of CO instabilities in the presence of preformed pairs. In fact, if COs coexist with coherent Cooper pairs in the SC state, they may also coexist with preformed pairs above $T_c$. In this context, the microscopic theory of high-$T_c$ superconductivity may not depend on either the one-gap or two-gap scenario being the sufficient condition.

Regardless of the theoretical views, a feasible empirical approach to verify the validity of the ``two-gap'' scenario is to conduct vortex-state quasiparticle tunneling spectroscopic studies. Vortices induced by applied magnetic fields in type-II superconductors are known to form periodic regions within which the SC order parameter is locally suppressed~\cite{Abrikosov57}, leading to continuous bound states~\cite{Caroli64} and an enhanced local density of states (LDOS) at zero energy (referenced to the Fermi level) inside vortices~\cite{Hess90}. Consequently, if a CO coexists with SC in the ground state, it should be revealed inside the vortex core as a gapped feature at a CO energy $V_{\rm CO}$ upon local suppression of SC. This behavior would be in sharp contrast to the enhanced zero-energy LDOS near the center of vortex cores had SC been the sole ground-state order parameter. For pure $d_{x^2-y^2}$-wave pairing symmetry, additional four-fold ``star-shape'' orientational dependent conductance is expected due to the existence of nodal quasiparticles~\cite{Franz98}.

Earlier studies of the vortex-state scanning tunneling spectroscopy (STS) on two hole-type cuprates $\rm YBa_2Cu_3O_{7-\delta}$ (Y-123)~\cite{Fischer07,Maggio-Aprile95} and $\rm Bi_2Sr_2CaCu_2O_{8+x}$ (Bi-2212)~\cite{Pan00,Hoffman02} have shown no evidence for either the zero-energy conductance peak or the star-shape conductance pattern inside the vortex cores. In Bi-2212 it was found that by integrating the excess field-induced spectral weight up to a finite energy 12 meV, the resulting spatially resolved conductance map revealed weak checkerboard-like modulations inside each vortex~\cite{Hoffman02}. These LDOS modulations have been attributed to the presence of a coexisting competing order (CO) such as CDW~\cite{Kivelson03} or SDW~\cite{Polkovnikov02} in Bi-2212 upon the suppression of superconductivity inside the vortices. Additionally, recent zero-field spatially resolved STS studies of a one-layer electron-type cuprate superconductor $\rm Pr_{0.88}LaCe_{0.12}CuO_4$ (PLCCO) have shown supporting evidence for a bosonic mode related to spin excitations in the SC state~\cite{Niestemski07}. However, to date there have not been any spatially resolved {\it vortex-state} STS studies reported on the electron-type cuprates.

Given various contrasts in the physical properties of electron- and hole-type cuprates, we expect the investigation of vortex-state quasiparticle tunneling spectra of electron-type cuprates to provide new insights into the feasibility of the CO scenario. In this letter, we provide the first detailed vortex-state investigation on an optimally doped electron-type cuprate $\rm Sr_{0.9}La_{0.1}CuO_2$ (La-112).

\section{Experimental}
The unit cell of La-112 is nearly cubic, with in-plane and c-axis lattice constants being 0.395 nm and 0.341 nm, respectively. The superconducting coherence length in the CuO$_2$ plane is $\xi _{ab} \sim 4.86$ nm and along c-axis is $\xi _c \sim 0.52$ nm, which is longer than the c-axis lattice constant~\cite{KimMS02}. Various bulk properties such as the anisotropic upper critical fields and irreversibility fields of this system have been characterized previously~\cite{Zapf05}.

The spatially resolved differential tunneling conductance ($dI/dV$) vs. energy ($\omega$) spectra for the quasiparticle LDOS maps were obtained with our homemade cryogenic scanning tunneling microscope (STM). The STM has a base temperature of 6 K, variable temperature range up to room temperature, magnetic field range up to 7 Tesla, and ultra-high vacuum capability down to a base pressure $< 10^{-9}$ Torr at 6 K. Given its highly three-dimensional chemical structure, the surface of La-112 cannot be cleaved as in the case of Bi-2212, and must be chemically etched~\cite{Vasquez94}. The chemically etched surface is found to be stoichiometric from x-ray photoemission spectroscopy (XPS)~\cite{Vasquez01}, which also yields high-quality and reproducible zero-field STS at cryogenic temperatures~\cite{ChenCT02}. For the STS data reported in this work, the tunnel junction resistance has always been kept at $\sim 1 \rm G \Omega$ to ensure high quality junctions.

For each constant temperature ($T$) and magnetic field ($H$), the experiments were conducted by tunneling currents along the c-axis of a single-crystal grain of La-112 under a range of bias voltages at a given location. The finite-field tunneling spectra were always taken under the zero-field-cool condition. Current ($I$) vs. voltage ($V$) measurements were repeated pixel-by-pixel over an extended area of the sample, and a typical two-dimensional map consisted of $(128 \times 128)$ pixels. Therefore, the spatial resolution was determined by the area of spectral measurements divided by the total number of pixels. To remove slight variations in the tunnel junction resistance from pixel to pixel, the differential conductance at each pixel is normalized to its high-energy background, as specified in Ref.~\cite{ChenCT02}.

\section{Results}
A representative zero-field tunneling spectrum of the normalized differential conductance ($dI/dV$) relative to the quasiparticle energy ($\omega = eV$, $V$ being the bias voltage) taken at 6 K is shown in Fig.~1(a), together with three different theoretical fitting curves to be discussed later. Here we denote the energy associated with the spectral peaks as $\Delta _{\rm eff}$. With increasing temperature, $\Delta _{\rm eff} (T)$ steadily decreases, as shown in the main panel of Fig.~1(b) and summarized in the inset of Fig.~1(b). Moreover, we find relatively homogeneous zero-field quasiparticle tunneling spectra, as manifested by the $\Delta _{\rm eff}$ histogram in Fig.~1(c). This finding is in stark contrast to the nano-scale spectral variations in optimal and underdoped Bi-2212~\cite{McElroy05}.

\begin{figure}
\includegraphics[width=3.45in]{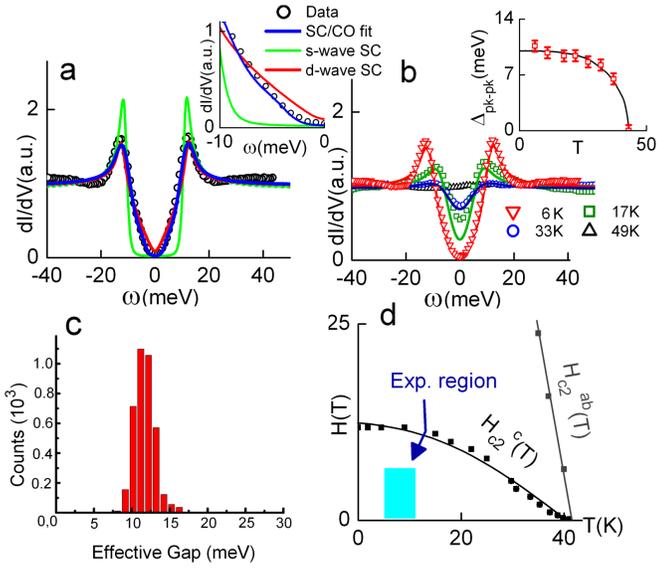}
\caption{Zero-field tunneling spectra and critical fields of La-112: (a) A $(dI/dV)$ vs. energy ($\omega$) tunneling spectrum (circles) at $T$ = 6 K normalized to its high-energy background~\cite{ChenCT02}, together with three different theoretical fitting curves that assume pure $s$-wave SC (green), pure $d_{x^2-y^2}$-wave SC (red), and coexisting $d_{x^2-y^2}$-wave SC and commensurate SDW (blue) with fitting parameters $\Delta _{\rm SC} = (12.0 \pm 0.2)$ meV, $V_{\rm CO} = V_{\rm SDW} = (8.0 \pm 0.2)$ meV, and SDW wave-vector $\textbf{Q} = (\pi, \pi)$. The inset shows the zoom-in comparison of the data and three different theoretical fittings. As elaborated in Discussion and detailed in Refs.~\cite{ChenCT07,Beyer08}, coexisting SC and SDW provides the best fitting, and the spectral peak is associated with $\Delta _{\rm eff}$. (b) Evolution of quasiparticle spectra with temperature from $T$ = 6 K to $T$ = 49 K $> T_c = 43$ K, showing absence of PG above $T_c$. The solid lines are theoretical fitting curves using temperature Green's function~\cite{ChenCT07,Beyer08} that yield the correct empirical $\Delta _{\rm eff} (T)$ given in the inset. The temperature-dependent ($\Delta _{\rm SC}$, $V_{\rm CO}$) values in units of meV are (12.0, 8.0) for 6 K, (9.5, 2.0) for  17 K, (7.5, 0) for 33 K, (0, 0) for 49 K. The zero-field $\Delta _{\rm eff}$-vs.-$T$ data (red circles) in the inset of Fig.~1(b) largely follow the BCS temperature dependence (solid line). (c) A histogram of the zero-field effective gap $\Delta _{\rm eff}$ at $T$ = 6 K, showing relatively homogeneous gap values $\Delta _{\rm eff} = 12.2 \pm 0.8$ meV over an area of $(64 \times 64)$ nm$^2$. (d) The experimental $(H,T)$ variables investigated in this work are shown in reference to the upper critical fields $H_{c2} ^{ab}$ for $H \perp \hat c$ and $H_{c2} ^c$ for $H \parallel \hat c$ of La-112~\cite{Zapf05}. Our experimental conditions are consistent with the $T \ll T_c$ and $H \ll H_{c2}$ limit, as shown in the shaded area.}
\label{fig.1}
\end{figure}

Upon applying magnetic fields $H$ with the condition $0 < H \ll H_{c2} ^{ab,c}$, the quasiparticle tunneling spectra exhibit strong spatial inhomogeneity, as exemplified in Fig.~2(a) for $H$ = 1 T over a $(64 \times 64)$ nm$^2$ area and Fig.~2(c) for $H$ = 2T over a $(65 \times 50)$ nm$^2$ area. Given that we are primarily interested in achieving high spatial resolution in order to investigate the inter- to intra-vortex spectral evolution, we first focus on smaller spatial maps in finite fields. Even for the smaller maps, vortices are still clearly visible at both 1T and 2T with  averaged vortex lattice constants $a_B =$ 52 nm and 35 nm, respectively, which are comparable to the theoretical value $a_B = 1.075 (\Phi _0 /B)^{1/2}$. Specifically, we identify the location of vortices by plotting the spatial map of the conductance power ratio, defined as the value of $(dI/dV)^2$ at the spectral peak energy $\omega = \Delta _{\rm eff}$ relative to that at the zero energy $\omega = 0$. We find that the presence of vortices is associated with the local minimum of the conductance power ratio, which is consistent with enhanced zero-energy quasiparticle density of states inside the vortex core. We further note that the average radius of the vortices is comparable to the superconducting coherence length $\xi _{ab}$. 

In addition to the high resolution vortex maps in Fig.~2(a) and Fig.~2(c), we illustrate a larger area vortex map for $H = 1.5 T$ with reduced spatial resolution in Fig.~2(e), which shows disordered vortices over a $(160 \times 152)$ nm$^2$ area. Despite the disorder, we find that the total flux is still conserved within the area studied. That is, the total number of vortices multiplied by the flux quantum is equal to the magnetic induction multiplied by the area, within experimental errors. We obtain an averaged vortex lattice constant $a_B = 42$ nm, comparable to the theoretical value of $a_B = 40$ nm. For comparison, we show in Fig.~2(f) the zero-field conductance power ratio map over the same area as in Fig.~2(e). The conductance ratio map at $H = 0$ clearly demonstrates significant contrast to the maps at finite fields, further verifying our observation of vortices in La-112. However, we note that the shape of all vortices observed in our experiments is generally irregular. Possible cause for the irregular vortex shape may be due to microscopic disroder in the sample as well as surface roughness after chemical etching. Additionally, possible interaction between the STM tip and vortices may have also contributed to the irregular vortex shape. 

\begin{figure}
\includegraphics[width=3.45in]{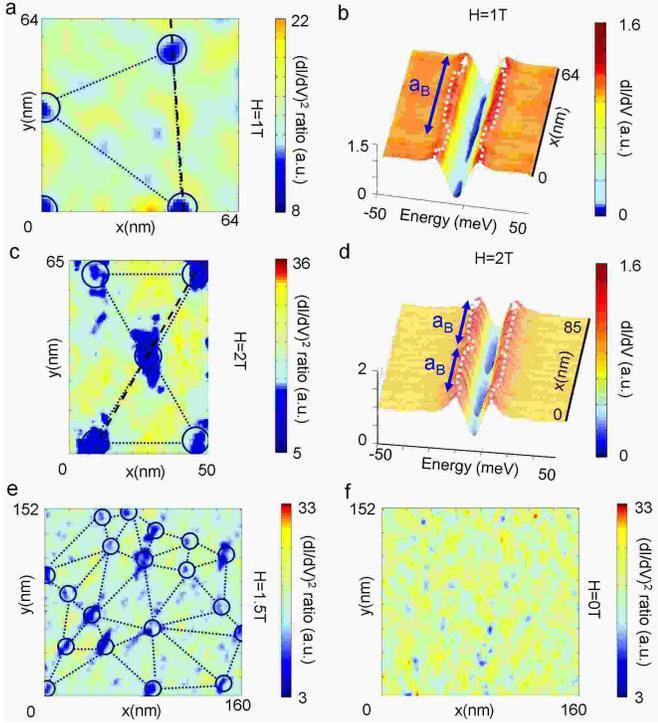}
\caption{Vortex maps of La-112 at $T$ = 6 K and for $H \parallel c$-axis: (a) A spatial map of the conductance power ratio (in log scale) taken over a $(64 \times 64)$ nm$^2$ area with $H$ = 1 T, showing a zoom-in view of vortices separated by an average vortex lattice constant $a_B = 52$ nm, which compares favorably with the theoretical value of 49 nm. The average radius of the vortices (indicated by the radius of the circles) is $(4.7 \pm 0.7)$ nm, comparable to the SC coherence length $\xi _{ab} = 4.9$ nm. Here the conductance power ratio is defined as the ratio of $(dI/dV)^2$ at $|\omega| = \Delta _{\rm eff}$ and that at $\omega = 0$. (b) Spatial evolution of the conductance $(dI/dV)$ along the black dashed line cutting through two vortices in (a) for $H$ = 1 T, showing significant modulations in the zero-bias conductance and slight modulations in the peak-to-peak energy gap. These modulations generally follow a periodicity of $a_B$, and the zero-bias conductance (the peak-to-peak gap value) reaches a maximum (minimum) inside the vortex core. (c) A spatial map of the conductance power ratio (in log scale) taken over a $(65 \times 50)$ nm$^2$ area with $H$ = 2 T, showing a zoom-in view of vortices with an average vortex lattice constant $a_B = 35$ nm, which is consistent with the theoretical value. The averaged vortex radius (indicated by the radius of the circles) is $(5.0 \pm 1.3)$ nm. (d) Spatial evolution of the conductance is shown along the black dashed line cutting through three vortices in (b) for $H$ = 2 T. (e) A spatial map of the conductance power ratio (in log scale) taken over a larger $(160 \times 152)$ nm$^2$ area with $H$ = 1.5 T, showing disordered vortices. Nonetheless, the average vortex lattice constant $a_B = 42$ nm remains consistent with Arbikosov's theory. (f) A spatial map of the conductance power ratio (in log scale) taken at $H = 0$ over the same $(160 \times 152)$ nm$^2$ area as in (e), showing a relatively homogeneous map of the conductance ratio in contrast to all other maps taken at $H > 0$.}
\label{fig.2}
\end{figure}

To better evaluate the spatial evolution of the tunneling spectra, we take a line cut through multiple vortices and illustrate the corresponding conductance $(dI/dV)$ spectra in Fig.~2(b) for $H$ = 1 T and in Fig.~2(d) for $H$ = 2 T. We find that the tunneling spectra inside vortices exhibit PG-like features rather than a peak at the zero bias, similar to the findings in hole-type cuprates of Y-123 and Bi-2212~\cite{Fischer07,Maggio-Aprile95,Pan00}. The PG energy is smaller than the zero-field $\Delta _{\rm eff} (H = 0)$ and comparable to the zero-field competing order energy $V_{\rm CO} (H = 0)$ derived from the Green's function analysis (see Discussion and Refs.~\cite{ChenCT07,Beyer08}). Both the zero-bias conductance and the peak-to-peak energy gap exhibit modulations with a periodicity of $a_B$; the zero-bias conductance shows a maximum inside vortices, whereas the peak-to-peak energy gap reaches a minimum inside vortices. Upon increasing magnetic field, the energy associated with the peak features at $\Delta_{\rm eff}(H)$ {\it outside} vortices decreases slightly and the linewidth of the peaks broadens, whereas the PG-energy at $V_{\rm CO}$ {\it inside} vortices remains constant, as shown in Figs.~3(a)-3(d) for comparison of representative inter- and intra-vortex spectra taken at $H$ = 1, 2, 3.5 and 6 T.

\begin{figure}
\includegraphics[width=3.45in]{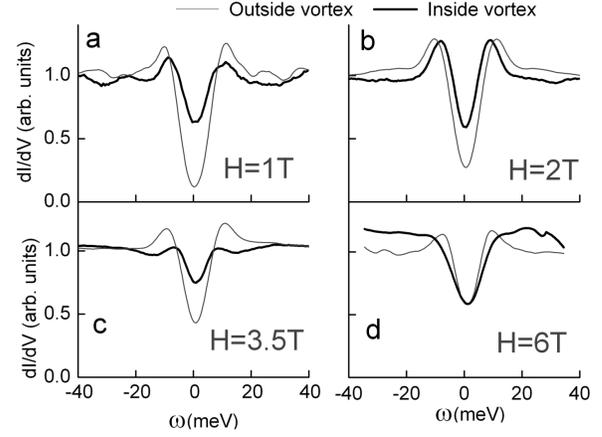}
\caption{Evolution of the inter- and intra-vortex quasiparticle tunneling spectra with magnetic field in La-112 for (a) $H$ = 1 T and $T$ = 6 K, (b) $H$ = 2 T and $T$ = 6 K, (c) $H$ = 3.5 T and $T$ = 6 K, (d) $H$ = 6 T and $T$ = 11 K, where the PG spectra at the center of vortex cores are given by the thick lines and those exterior to vortices are given by the thin lines. We note that the peak features associated with the inter-vortex spectra broaden with increasing $H$, and the zero-bias conductance of the inter-vortex spectra increases with increasing $H$.}
\label{fig.3}
\end{figure}

To fully account for the statistics of the spectral evolution with increasing fields, we illustrate the histograms of the characteristic energies identified from all spectra taken over spatial maps varying from $(50 \times 50)$ nm$^2$ to $(100 \times 100)$ nm$^2$ in Fig.~4(b). The characteristic energy of each spectrum is defined as one half of the peak-to-peak energy $\Delta _{\rm pk-pk}$. We find that each histogram can be fit by a Lorentzian functional form, and we associate the peak energy with $\Delta _{\rm eff} (H)$, which decreases slightly with increasing magnetic field, as summarized in Fig.~4(a). Additionally, there is an apparent low energy ``cutoff'' at $V_{\rm CO} = (8.5 \pm 0.6)$ meV for all histograms. Following the analysis described in Discussion, we may attribute the decrease in $\Delta _{\rm eff} (H)$ to that in $\Delta _{\rm SC} (H)$, as shown in Fig.~4(a). This behavior is in stark contrast to the histogram expected for a conventional type-II superconductor in a magnetic field $H$ if SC were the sole order parameter. In the latter situation and for $H \ll H_{c2}$ and $T \ll T_c$, a fraction of the total spectral weight on the order of $(\pi \xi _{ab} ^2/2)/(\sqrt{3} a_B ^2/4)$ in the histogram would have been downshifted to energies $\omega < \Delta _{\rm SC} (0)$ due to the suppression of SC inside vortices. The downshifted spectral weight should have been concentrated at $\omega = 0$, as schematically illustrated in Fig.~4(c). The zero-bias spectral weight would have been linearly proportional to $H$ and become readily observable in our experiments, e.g. $\sim 25 \%$ for $H$ = 6 T.

\begin{figure}
\includegraphics[width=2.4in]{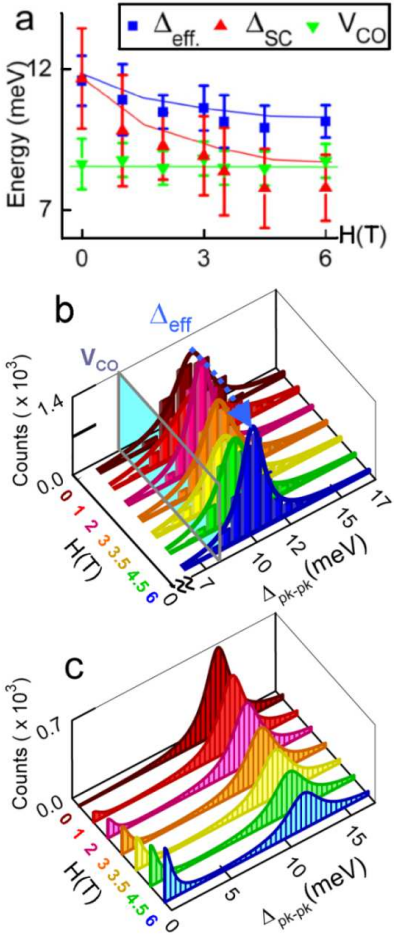}
\caption{Quasiparticle spectral evolution in La-112 as a function of magnetic field ($H$): (a) Magnetic field dependence of the characteristic energies $\Delta _{\rm eff}$, $\Delta _{\rm SC}$ and $V_{\rm CO}$ for $H$ = 1, 2, 3, 3.5, 4.5 and 6 T. For each field $H$ the corresponding energy histogram is obtained by identifying one-half of the quasiparticle spectral peak-to-peak energy separation as $\Delta _{\rm pk-pk}$ over $(50 \times 50)$ nm$^2$ to $(100 \times 100)$ nm$^2$ areas. Each histogram can be fit with a Lorentzian functional form, and the peak position of the Lorentian is identified as $\Delta _{\rm eff} (H)$. The low-energy cutoff of the histogram is identified as $V_{\rm CO}$, and the SC gap for a given field is defined as $\Delta _{\rm SC} (H) = \lbrace \lbrack \Delta _{\rm eff} ^2 (H) - V_{\rm CO} ^2  \rbrack ^{1/2} / \cos (2 \phi _{\rm AF}) \rbrace$, where $\phi _{\rm AF} \approx 25 ^{\circ}$ is an angle associated with the antiferromagnetic ``hot spots''~\cite{Matsui05}. Empirically, we find that $V_{\rm CO}$ is nearly constant, whereas $\Delta _{\rm SC} (H)$ decreases slightly with increasing $H$. (b) Energy histograms of La-112 determined from our quasiparticle tunneling spectra of La-112, showing the spectral evolution with $H$. Note that there is no zero-bias conductance peak in the vortex-state, and that a low-energy cutoff at nearly a constant value $V_{\rm CO} = (8.5 \pm 0.6)$ meV exists for all fields. (c) Schematic illustration of the theoretical histograms in a conventional type-II superconductor under the conditions $T \ll T_c$ and $H \ll H_{c2}$. With increasing magnetic field, the spectral weight in the vortex state of a conventional type-II superconductor shifts to lower energies and peaks at $\omega = 0$. The downshifted spectral weight for a given field $H$ relative to the total spectral weight in $H = 0$ is approximated by the ratio of the vortex core area relative to the Abrikosov vortex unit cell, $(\pi \xi _{ab} ^2/2)/(\sqrt{3} a_B ^2/4)$.}
\label{fig.4}
\end{figure}

\section{Discussion}
The complete absence of a zero bias conductance peak in the vortex-state quasiparticle spectra of La-112, the occurrence of PG-like behavior revealed inside the vortex core, and the existence of a low-energy cutoff at $V_{\rm CO}$, cannot be easily explained by the scenario of pure SC in the ground state. On the other hand, these anomalous experimental findings may be compared with the scenario of coexisting CO and SC by means of the Green's function analysis detailed in Refs.~\cite{ChenCT07,Beyer08} together with realistic bandstructures~\cite{Matsui05}. In Fig.~1(a), we compare three different theoretical fitting curves with the zero-field tunneling spectrum of La-112 at 6 K. The fittings assume pure $s$-wave SC (green), pure $d_{x^2-y^2}$-wave SC (red), and coexisting $d_{x^2-y^2}$-wave SC and commensurate SDW (blue) with the following fitting parameters: SC gap $\Delta _{\rm SC} = (12.0 \pm 0.2)$ meV, CO energy $V_{\rm CO} = (8.0 \pm 0.2)$ meV, and CO wavevector $\textbf{Q} = (\pi, \pi)$ for the SDW. Here we note that the consideration of commensurate SDW~\cite{Schrieffer89} as the relevant CO is consistent with neutron scattering data from electron-type cuprate superconductors~\cite{Motoyama06}. As shown in Refs.~\cite{ChenCT07,Beyer08}, the coexisting SC and CO scenario yields only one set of spectral peaks at $\omega = \pm \Delta _{\rm eff}$ as long as the condition $V_{\rm CO} \le \Delta _{\rm SC}$ is satisfied, and $\Delta _{\rm eff} \equiv \lbrack \Delta _{\rm SC} ^2 + V_{\rm CO} ^2 \rbrack ^{1/2}$~\cite{ChenCT07,Beyer08}. We further note that attempts to fit the spectra with the Dynes model~\cite{Dynes78} that assumes pure SC together with extra quasiparticle lifetime-broadening would have led to substantial increase in the zero-energy LDOS, inconsistent with the empirical observation of vanishing zero-field LDOS at $\omega \sim 0$.

The analysis shown in Fig.~1(a) suggests that the CO scenario best accounts for the zero-field spectral details at $T \ll T_c$. Interestingly, the CO energy derived from the zero-field analysis ($V_{\rm CO} = (8.0 \pm 0.2)$ meV) is consistent with the PG energy observed inside the vortex core, implying that the intra-vortex PG has the same physical origin as the zero-field CO. Additionally, the CO scenario can account for the temperature dependence of the zero-field tunneling spectra, as shown in the main panel of Fig.~1(b), where the theoretical fitting curves (lines) to the experimental data (symbols) are obtained by using the temperature Green's function and temperature-dependent $\Delta _{\rm SC}$ and $V_{\rm CO}$ values that yield the correct empirical $\Delta _{\rm eff} (T)$ given in the inset of Fig.~1(b)~\cite{ChenCT07,Beyer08}. Given that neither the intra-vortex PG nor the spectral evolution of $\Delta _{\rm eff}$ with magnetic field may be explained by pure $d_{x^2-y^2}$-wave SC, we find that the CO scenario with the relevant CO being a commensurate SDW provides better account for our experimental findings in the optimally doped La-112 system. 

As an interesting comparison, our recent spatially resolved vortex-state quasiparticle tunneling studies on a hole-type optimally doped cuprate $\rm YBa_2Cu_3O_{7-\delta}$ (Y-123) also revealed PG-like features and field-induced LDOS modulations inside vortices~\cite{Beyer08a}, except that the PG energy in Y-123 is {\it larger} than $\Delta _{\rm SC}$. This finding may also be interpreted as a CO being revealed upon the suppression of SC. Furthermore, the larger energy associated with the PG-like features inside vortices of Y-123 is consistent with the presence of a zero-field PG temperature {\it higher} than $T_c$. These findings from the vortex-state quasiparticle spectra of Y-123 are in contrast to those of La-112; in the latter no zero-field PG exists above $T_c$ and the field-induced PG energy is {\it smaller} than $\Delta _{\rm SC}$.   

However, it is worth discussing two alternative scenarios for the absence of zero-bias conductance peak inside vortices. First, in the case of $s$-wave SC with parabolic bandstructures, it has been shown that the bound states inside the vortex core can lead to a zero-bias conductance peak everywhere inside the vortex core under finite thermal smearing~\cite{Shore89}. On the other hand, in the limit of small thermal smearing, it is found numerically~\cite{Shore89} that for bound states of high angular momemta, the LDOS at $0 < r < \xi _{\rm SC}$ inside the vortex core ($r$: the distance from the center of a vortex core) may acquire a dip-like feature in the conductance~\cite{Shore89} rather than a peak at zero bias, which seem to explain our experimental findings. Nonetheless, for pure $d_{x^2-y^2}$-wave superconductors such as the cuprates, no true bound states exist inside the vortex core~\cite{Franz98}. Moreover, in contrast to the theoretical predictions of a zero-bias conductance peak at $r \approx 0$ and PG-like features at continuously varying energies $0 < \omega < \Delta _{\rm SC}$ for $0 < r < \xi _{\rm SC}$ and $T \ll T_c$~\cite{Shore89}, our spectral studies revealed a nearly constant PG energy $(8.5 \pm 0.6)$ meV everywhere inside the vortex core, with approximate $(0.5 \times 0.5) \rm nm^2$ spatial resolution for a vortex core radius $\sim 5$ nm. Therefore, the field-induced PG-like features in La-112 cannot be easily explained by the scenario of high-angular-momentum bound states in an $s$-wave superconductor. Additionally, our observation of a field-induced PG energy larger than $\Delta _{\rm SC}$ in Y-123~\cite{Beyer08a} cannot be reconciled with the notion of bound states inside the vortex core of a pure superconductor, because the bound-state energy inside a vortex core of a pure SC system cannot exceed $\Delta _{\rm SC}$. 

Second, recent theoretical studies~\cite{Balents05,Nikolic06} have demonstrated that the inclusion of quantum fluctuations for a single vortex in a pure $d_{x^2-y^2}$ superconductor can give rise to suppression of the zero-energy LDOS peak at the vortex center, which corroborates our previous empirical findings of strong field-induced quantum fluctuations in cuprate superconductors~\cite{Beyer07}. Nonetheless, quantitative discrepancies remain between experimental results and theory that considers quantum fluctuations alone for a single vortex without including CO's and vortex-vortex correlations~\cite{Balents05,Nikolic06}. Further investigation appears necessary to fully account for the experimental observation. 

Finally, we attribute the difficulties previously encountered with directly identifying vortices in electron-type cuprate superconductors to the presence of a PG feature inside the vortex core and the decreasing contrast between the intra- and inter-vortex spectra with increasing magnetic field, as exemplified in Figs.~3(a)--(d). In particular, the SC gap values of other electron-type cuprates are much smaller than our La-112 system, rendering the task of distinguishing inter-vortex spectra from intra-vortex spectra even more difficult. We further note that our finding of a hiddent PG inside vortex cores in the infinite-layer La-112 system is consistent with a previous bulk break-junction study of one-layer electron-type cuprate superconductors $\rm Pr_{2-x}Ce_xCuO_{4-y}$ and $\rm Pr_{2-x}Ce_xCuO_{4-y}$~\cite{Alff03}, where a spatially averaged field-induced PG with an energy smaller than the SC gap has been observed for a range of doping levels. 

\section{Conclusion}
We have demonstrated in this letter the first-time observation of vortices in an electron-doped cuprate superconductor. We have also revealed from spatially resolved quasiparticle tunneling spectroscopy a hidden pseudogap inside vortices and unconventional spectral evolution with temperature and magnetic field. None of these results can be easily explained by the scenario of pure superconductivity in the ground state of the cuprates, thereby imposing important constraints on the theory of high-$T_c$ superconductivity.

\acknowledgments
The work at Caltech was jointly supported by the Moore Foundation and the Kavli Foundation through the Kavli Nanoscience Institute at Caltech, and the NSF Grant DMR-0405088. The work at Sogang University was supported by the Center of Superconductivity from the program of Acceleration Research of MOST/KOSEF of Korea and Special fund of Sogang University. ADB acknowledges the support of Intel Graduate Fellowship.

\end{document}